# Spectroscopic Evidence for the Superconductivity of Elemental Metal Y under Pressure

Zi-Yu Cao,[1,*] Harim Jang,[1,*] Seokmin Choi,[1] Jihyun Kim,[1,2] Suyoung Kim,[1] Jian-Bo Zhang,[3] Anir S. Sharbirin,[4] Jeongyong Kim,[4] and Tuson Park[1,†]

[1] *Center for Quantum Materials and Superconductivity (CQMS) and Department of Physics, Sungkyunkwan University, Suwon 16419, Republic of Korea*

[2] *Institute of Basic Science, Sungkyunkwan University, Suwon 16419, Republic of Korea*

[3] *Center for High-Pressure Science & Technology Advanced Research, Beijing, 100094, People's Republic of China*

[4] *Department of Energy Science, Sungkyunkwan University, Suwon 16419, Republic of Korea*

\* These authors contributed equally to this work.

† Correspondence should be addressed to TP (tp8701@skku.edu)

**Very high applied pressure induces superconductivity with the transition temperature ($T_c$) exceeding 19 K in elemental yttrium, but relatively little is known about the nature of that superconductivity. From point-contact spectroscopy (PCS) measurements in a diamond anvil cell (DAC), a strong enhancement in the differential conductance is revealed near the zero-biased** **voltage owing to Andreev reflection, a hallmark of the superconducting (SC) phase. Analysis of the PCS spectra based on the extended Blonder-Tinkham-Klapwijk (BTK) model indicates two SC gaps at 48.6 GPa, where the large gap $\Delta_L$ is 3.63 meV and the small gap $\Delta_S$ is 0.46 meV. When scaled against a reduced temperature, both small and large SC gaps collapse on a single curve that follows the prediction from BCS theory. The SC gap-to-$T_c$ ratio is 8.2 for the larger gap, and the initial slope of the upper critical field is −1.9 T/K, indicating that Y belongs to a family of strongly coupled BCS superconductors. The successful application of PCS to Y in DAC environments demonstrates its utility for future research on other pressure-induced high-$T_c$ superconductors.**

Superconductivity in elemental metals has been a cornerstone in discovering superconducting materials at elevated temperatures [1, 2, 3]. In simple metals where their electronic properties can be reasonably approximated by the free-electron model, superconductivity is typically suppressed under physical pressure owing to its tendency to lower the density of states at the Fermi level [3]. Among those elements that do not superconduct at ambient pressure, nonmagnetic d-orbital elements such as Sc, Y, and Lu are exceptional in that the pressure-induced SC state is dramatically enhanced with increasing pressure, exhibiting complexity under extreme compression [4, 5]. The importance of understanding SC properties in the elemental superconductors has been noted by the recent discoveries of an SC phase in the metal hydrides $MH_n$ (M=La, Y), where a superconducting transition temperature ($T_c$) is observed at temperatures above 200 K at pressures of several megabars [6, 7, 8, 9].

Yttrium (Y), which participates in the superhydride superconductor $YH_n$ with a $T_c$ of 243 K [8, 9], exhibits superconductivity under pressure at temperatures as high as 20 K, one of the highest $T_c$s among the elemental superconductors. At an external pressure of 11 GPa, Y transforms into the superconducting state below 1.2 K [10]. AC magnetic susceptibility and electrical resistivity measurements show that $T_c$ monotonically increases to 20 K at 100 GPa and decreases at higher pressures, showing a peak near 100 GPa, where there is a structural change [5, 11]. Although the $T_c(P)$ phase diagram of yttrium and its structural evolution with decreasing volume have been established [12, 13], SC properties such as the upper critical field and SC gap, which are important for understanding the mechanism of its superconductivity, have yet to be studied.

To address these issues, we have carried out point-contact spectroscopy (PCS) measurements of Y metal in a diamond anvil cell (DAC). The PCS measurements, which were performed as a function of both temperature and magnetic field at 48.6 GPa, reveal that the differential conductance ($dI/dV$) of Y is best described by an $s$-wave superconducting order parameter with two SC gaps, i.e., $\Delta_L(0) = 3.63$ meV and $\Delta_S(0) = 0.46$ meV. The SC gap-to-$T_c$ ratio, $2\Delta_L(0)/k_B T_c$, for the larger gap is 8.2, which is much higher than the 3.53 expected for a weakly coupled Bardeen–Cooper–Schrieffer (BCS) superconductor; instead, it is comparable to that of strongly correlated superconductors such as high-$T_c$ cuprates, heavy fermions, and Fe-based compounds [14]. In support of the unusual superconductivity, the initial slope of the upper critical field at $T_c$ is −1.9 T/K at 48.6 GPa, which is ten times that of the two-gap superconductor $MgB_2$ [15] and larger than that of the Fe-based superconductor LiFeAs [16]. The successful development of the PCS technique in a DAC not only reveals the strongly coupled superconductivity of elemental Y but is also expected to provide a much-needed method to guide efforts to understand the SC properties of high-$T_c$ superconductors under extreme environments [7].

The electrical resistivity of Y under pressure is plotted as a function of temperature in Fig. 1a. The resistivity at 21.1 GPa decreases with decreasing temperature, exhibiting metallic behavior. However, a signature of the SC phase transition is absent at temperatures above 6 K. Increasing the pressure to 35.4 GPa causes the resistivity to drop sharply to zero at 9.9 K owing to the SC transition. A further increase in pressure gradually enhances $T_c$ to 19.1 K at 90.2 GPa, above which $T_c$ decreases slightly with pressure. Figure 1b displays the dependence of $T_c$ on pressure, where the results obtained in this work are represented by star symbols that track data obtained from previous ac magnetic susceptibility (30–125 GPa) and resistivity measurements (11–17 and 80–166 GPa) [5, 10, 11, 12]. Even though the $T_c$ obtained in this work is slightly higher than that in previous studies, the pressure dependence of $T_c$ and the positions of the peak near 100 GPa are similar.

The change in the pressure-induced SC state of Y in the presence of a magnetic field at 33.4 and 48.6 GPa are displayed in Fig. 2a and 2b, respectively. The resistivity reveals a peak near $T_c$ and decreases to zero for both pressures owing to the SC phase transition. A hump-like feature near $T_c$ is often observed in disordered superconductors, which can be ascribed to the development of SC puddles surrounded by normal state regions near $T_c$ [8, 17, 18]. At 9 T, which is the strongest magnetic field available in this work, the SC phase at 48.6 GPa is still robust, and the onset of $T_c$ occurs at 3.9 K. The upper critical field ($H_{c2}$) observed in Y under pressure is the highest among bulk element superconductors, suggesting that the pressurized Y metal is a type-II superconductor [17, 19].

The magnetic field dependence of $T_c$ determined as the onset of the SC phase transition is plotted in Fig. 2c and 2d, where the initial slope (dashed line) of the upper critical field at $T_c$, $d(\mu_0 H_{c2})/dT$, is −1.2 and −1.9 T/K at 33.5 and 48.6 GPa, respectively. The upper critical field at absolute zero temperature of Y estimated from the Werthamer-Helfand-Hohenberg (WHH) theory for a dirty-limit superconductor, $\mu_0 H_{c2}(0) = -0.693 \cdot d(\mu_0 H_{c2})/dT|_{T=T_c} \cdot T_c$, is 6.5 and 13.6 T at 33.4 and 48.6 GPa, respectively [20]. The SC coherence length ($\xi$) derived from the relation of $\mu_0 H_{c2}(0) = \Phi_0/2\pi\xi(0)^2$ is 7.0 nm and 4.8 nm at 33.4 GPa and 48.6 GPa, respectively, where $\Phi_0$ is the flux quantum. A kink-like feature in $\mu_0 H_{c2}(T)$ near 4.0

K at 33.4 GPa and 5.5 K at 48.6 GPa, indicated by the upper arrows, is a deviation from the prediction for a single-band scheme [20, 21].

Establishing point-contact junctions that could provide appropriate energy-resolved spectroscopic information, i.e., negligible energy dissipation at the junction, in the presence of external pressure is essential for unveiling the SC gap structure of Y. Even though there has been a plethora of efforts in the implementation of the PCS technique under pressure, they are limited to low-pressure ranges below ~3 GPa in clamp-type cells [22, 23]. Here, we successfully applied the PCS technique in a DAC environment up to 48.6 GPa, providing an essential breakthrough in probing low-energy physics under extreme conditions surpassing the existing pressure limit. Figure 3(a) illustrates the concept of the Pt/Y point-contact junction formed inside the DAC, where we attached the edge of a sharp Pt cut to the sample surface by mechanical pressure (see Methods for more details). Detailed diagnostics on the junction in the DAC environment indicate that the probed spectra reflect the SC properties of Y (see Section B in Supplementary Information (SI)).

Figure 3 shows the dependence of the differential conductance $dI/dV$ on the bias voltage at 48.6 GPa, which is obtained from the PCS. The signature of the Andreev reflection owing to the presence of an SC gap is representatively shown in Fig. 3b, where the broad peak in $dI/dV$ is overlaid with a small peak near zero-bias voltage [24, 25]. The solid and dashed lines are best fits based on the Blonder-Tinkham-Klapwijk (BTK) model for single and two $s$-wave SC gaps, respectively, showing that two-gap superconductivity is realized in elemental metallic Y. To take into account the dip feature observed near 10 mV, an intergrain Josephson effect (IGJE) that could lead to a dip feature at the edge of the Andreev reflection is introduced to the modified BTK model [26, 27]:

$$\frac{dI}{dV}(V) = G_0 \left( \frac{dV_{BTK}}{dI} + w_I \frac{dV_{IGJE}}{dI} \right)^{-1} \quad (1)$$

where $G_0$ is the differential conductance in the normal state and $w_I$ is the IGJE weight. The first term of Equation (1) corresponds to the modified BTK formula for the two-band *s*-wave SC pairing symmetry (denoted as bands L and S), which is expressed as follows:

$$\left.\frac{dI}{dV}\right|_{BTK} = w_B \left.\frac{dI}{dV}\right|_{band\,L} + (1-w_B) \left.\frac{dI}{dV}\right|_{band\,S} \quad (2)$$

Here, the contribution from each band is evaluated using the modified BTK formula [25]. The second term of Equation (1) is a contribution from the IGJE, which is the solution of the Fokker-Planck partial differential equation for a resistively shunted junction model with current fluctuations caused by thermal noise in the small capacitance limit [26]. The detailed numerical calculation procedures of Equations 1 and 2 are provided in section A in the SI.

The dependence on temperature of $dI/dV$ divided by its normal-state value at 10.5 K, $(dI/dV)/(dI/dV)_{10.5\,K}$, is selectively displayed in Fig. 3c with an offset for clarity. With increasing temperature, the broad peak from the Andreev reflection is gradually suppressed and disappears at temperatures above $T_c$ of 10.3 K (see Fig. S6 for details). Figure 3e is a color contour plot of the normalized conductance on the *T-V* axes, where green (red) represents larger (smaller) values. The suppressed IGJE regime in dark yellow surrounds the SC phase. The dependence on the magnetic field of the spectroscopic feature of Y, $(dI/dV)/(dI/dV)_{8.5\,T}$, at 5.0 K is summarized in Fig. 3d. The Andreev reflection is gradually suppressed with the magnetic field as in temperature and suppresses entirely above the critical field of 7 Tesla. As shown in the color contour plot on the *H-V* axes in Fig. 3f, however, the magnetic field suppresses the SC energy gap faster than the temperature because it suppresses not only the size of the SC energy gap but also $T_c$.

The dashed lines in Fig. 3c and 3d are least-squares fits of data to the modified BTK+IGJE model with contributions from the two SC gaps and IGJE terms, where contributions from the larger band $w_L$ and IGJE term $w_I$ are fixed to 0.75 and 0.7 over the whole temperature and field range, respectively (see Fig. S10 for other parameters obtained from the best fits). We note that the local heating effect of a nonballistic junction may develop a dip feature at the edge of the Andreev reflection. However, a systematic comparison between the anomalous dip structure in d$I$/d$V$ at a high biased voltage and the bulk critical current defined from the $I$–$V$ characteristic curve suggests that the dip feature near the SC gap is less likely from the local heating effect (see Fig. S4 and S5 in SI for details).

Multiple SC gaps have often been reported in novel SC compounds with short coherence lengths, such as high-$T_c$ cuprates, heavy fermion compounds, and other strongly correlated systems [28]. In the high-$T_c$ phase of sulfur hydrides, where $T_c$ reaches 203 K at 150 GPa, a two-gap model is predicted, where the small gap is critical for enhancing superconductivity [29]. The observation of two SC gaps in Y may be attributed to the $s$-$d$ transfer because the increase in electronic density of states in the $d$-band can change the electronic band structure at high pressures [30, 31]. Since the fractional volume of the ion core is relatively smaller under pressure, the occupation of the $d$ band (=$n_d$) at the Fermi energy is predicted to increase, thus introducing an SC phase in elemental metal Y [4, 5, 32, 33]. Identification of small pockets via electronic band structure calculations under high pressure will be necessary to understand the mechanism of multiband superconductivity in Y.

Figure 4a illustrates the temperature dependence of the SC energy gaps for two different point contacts P1 and P2 of Y at 48.6 GPa. The SC energy gaps $\Delta_L$ (band L) and $\Delta_S$ (band S) are similar for both P1 (open symbols) and P2 (solid symbols). The solid lines are best fits obtained from theoretical calculations based on the quasiclassical Eilenberger weak-coupling formalism with two isotropic SC gaps, where $\Delta_L(0)$ and $\Delta_S(0)$ are 3.63 and 0.45 meV, respectively (see Section D in SI for details) [34]. We note that the ab initio calculations on the temperature-dependent SC energy gaps of Y showed a strong coupling of SC pairing under pressure, which predicted $\Delta(0)$ = 1.34 and 1.64 meV at 26 and 31 GPa, respectively [35]. When plotted together with this work, as shown in Fig. S12 in the SI, the calculated values lie in the same line with a slope of ~0.1 meV/GPa, showing good agreement between the experimental results and theory. The magnetic field dependence of the SC energy gaps at $T$ = 5.0 K is shown in Fig. 4b for P1. The reduced gap versus reduced field ($H/H_{c2}$) follows the power law, $\Delta(H, 5.0\ \text{K}) = \Delta(0, 5.0\ \text{K}) \cdot (1 - H/H_{c2})^{1/2}$ (short-dashed line), which is often observed in the field dependence of a full-gap BCS superconductor [36]. We note that the Andreev reflection signature of the small SC gap $\Delta_S$ is smeared out and merged into the broad peak at high temperatures and high magnetic fields. Taken together with the reasonable description by BCS theory, however, the scaling of the small and large reduced gaps ($\Delta(T)/\Delta(0)$) against the reduced temperature ($T/T_c$) as well as the reduced magnetic field ($H/H_{c2}$) supports that the two SC gaps are coupled.

The ratio of the SC gap-to-$T_c$, $2\Delta(0)/k_B T_c$, serves as a criterion for the strength of the SC coupling constant relative to the BCS value of 3.53 for weak-coupling conventional superconductors. This ratio is higher for unconventional superconductors such as high-$T_c$ cuprates, heavy-fermion superconductors, and Fe-based superconductors [14]. In general, the gap ratio in multigap superconductors also deviates from the BCS value, with the ratio being above and below the weak-coupling limit for large and small SC gaps, respectively [37, 38]. For example, it is 4.4 and 1.9, respectively, for optimally doped Ba(Fe$_{1-x}$Co$_x$)$_2$As$_2$ [38]. Similar to other multigap superconductors, the SC gap ratio of Y at 48.6 GPa ($T_c$ = 10.3 K) is 8.2 and 1.0 for the large and small gaps, respectively. The deviation of the gap ratio from the BCS value indicates that the pressure-induced superconductivity of Y arises from strongly coupled electron-boson interactions, which is consistent with the theoretical calculations that predicted a strong coupling of SC pairing in Y [35, 39].

In conclusion, we presented spectroscopic evidence for the pressure-induced superconductivity of elemental metallic Y in a diamond anvil cell environment for the first time. The differential conductance of Y at 48.6 GPa is strongly enhanced near the zero-biased voltage owing to Andreev reflection, the hallmark of the SC state. The modified BTK model indicates two superconducting gaps of 3.63 and 0.46 meV, corroborated by a kink feature in the temperature dependence of the upper critical field. The large SC gap-to-$T_c$ ratio of 8.2 as well as the large initial slope of the upper critical field of 1.9 T/K suggest that Y belongs to a family of strongly coupled superconductors. The successful application of spectroscopic techniques under DAC environments is expected to open a new opportunity to probe the SC properties of high-$T_c$ superconductors under extreme conditions, including the recently reported superhydride high-$T_c$ superconductors.

**Methods**

**Pressure cell preparation.** A symmetric diamond anvil cell (DAC) with anvils in a 100 $\mu$m culet was used for resistance measurements in the 100 GPa pressure range, and a miniature Be-Cu DAC with anvils in a 300 $\mu$m culet was used for the upper critical field and point-contact spectroscopy measurements. A sample chamber with a diameter of 1/3 of the culet was drilled in a $c$-BN gasket (thickness of 20 $\mu$m), and salt was used as the pressure-transmitting medium. The fact that the resistance drop at $T_c$ is very sharp (e.g., $\Delta T_c \approx 0.3$ K and $\Delta T_c/T_c \approx 0.02$ at 45.4 GPa), as shown in Fig. 1(a), indicates that the pressure is close to the quasihydrostatic conditions. The sample was loaded in an argon-filled glove box, where four Pt slices adhered to a small piece of polycrystalline Y (Kojundo, 99.9%, thickness 2-3 $\mu$m) by mechanical pressure. The tip of the Pt was flattened to less than 1 $\mu$m in advance to avoid further broadening under pressure (see the inset of Fig. 1b). The pressure was measured by the position of the high-frequency diamond Raman signal in the symmetric DAC and the spectral shift of the fluorescence R1 peak of ruby in the Be-Cu DAC at room temperature after each measurement.

**Transport and spectroscopy measurements.** The van der Pauw configuration was used to measure the electrical resistance in a helium-4 closed-cycle refrigerator with a Lakeshore cryotronics 370 AC resistance bridge. The measurements of the upper critical field and point-contact spectroscopy under pressure were performed using a commercial cryostat PPMS model-6000 (9 T, Quantum Design).

With careful treatment of the Pt slices, the point-contact radius at the Pt/Y interface in the DAC can be reduced to below 5 $\mu$m, which is smaller than the extremity of the tip in the traditional method. The voltage responses to the DC current were measured at three small current steps ($\Delta I$), and the slope ($\Delta I/\Delta V$) was calculated to obtain the differential conductance value. The corresponding biased voltage in the junction is determined as an average measured voltage of three values, which is the x-axis value in the spectrum. The $\Delta I/\Delta V$ can be approximated to $dI/dV$ when the $\Delta I(\Delta V)$ is small enough, where all our measurements were performed with $\Delta I$ less than 0.5% of the whole curve. All values of differential conductance were obtained from the 3-point moving average to minimize the effects of the electromotive force in the circuit. The validity of the junction and its mathematical model for analysis are discussed in the Supplementary Information.

**Data availability**

All data that support the findings of this study are available from the corresponding author on reasonable request.

## Acknowledgments

We thank J. D. Thompson for the fruitful discussion. This work was supported by the National Research Foundation (NRF) of Korea through a grant funded by the Korean Ministry of Science and ICT (No. 2021R1A2C2010925).

## Author contributions

T.P. conceived and led the study. Z.Y.C. and S.C. performed electrical transport measurements at high pressures and calibrated pressures with the assistance of J.B.Z., A.S.S., and J.Y.K.; H.J. and Z.Y.C. carried out the PCS measurements with the assistance of S. K.; J.H.K. provided the sample. H.J., Z.Y.C., and T.P. analyzed the data and wrote the manuscript. All authors discussed the results and commented on the manuscript.

## Competing interests

The authors declare no competing interests.

**Figure Captions**

**Fig. 1: Electrical resistivity and *P-T* phase diagram of Y at different pressures.**

**a,** Pressure–temperature evolution of the resistivity of Y. The inset displays the resistivity over the entire temperature range (6–300 K) at which measurements were conducted at various pressures up to 109.4 GPa. **b,** *P-T* phase diagram of Y obtained from resistivity (circle symbols) and ac susceptibility (square symbols) measurements under pressure. The star symbols refer to our resistivity data in the symmetric cell (blue) and miniature Be-Cu cell (red). The resistivity measurements of $T_c$ from references [10, 12] are represented as navy and olive circles, respectively. $T_c$ collected from ac susceptibility measurements is indicated by square symbols [5, 11]. The dashed lines indicate the structural phase boundaries measured at room temperature [13]. As the pressure increases, the phase transition sequence for the crystal structure is hcp – α-Sm type – dhcp – dfcc – o$F$16. The inset shows a photograph of the van der Pauw configuration at 109.4 GPa.

**Fig. 2: Upper critical fields of Y under pressure.**

**a,** Electrical resistance of Y as a function of temperature for representative magnetic fields at 33.4 GPa and **b,** at 48.6 GPa. **c,** Temperature dependence of the upper critical fields, $\mu_0H_{c2}(T)$, at 33.4 GPa and **d,** 48.6 GPa, where $T_c$ is determined to be the temperature at which the resistance deviates from normal state behavior. We note that the $T_c$ onset is consistent with that determined from the PCS (see Fig. S4 in the Supplementary Information). The dashed lines indicate the initial slope of $\mu_0H_{c2}(T)$ near $T_c$. The arrows mark a kink in $\mu_0H_{c2}(T)$.

**Fig. 3: Quasiparticle scattering spectra at 48.6 GPa.**

**a,** Schematic diagram of point-contact spectroscopy on Y inside the diamond anvil cell. **b,** Normalized differential conductance, $(dI/dV)/(dI/dV)_{10.5\,K}$, at 2.0 K as a function of the bias voltage. The solid and dashed-dotted lines are the best fits to the two-band $s$-wave and single-band $s$-wave models, respectively. **c,** Temperature evolution of $(dI/dV)/(dI/dV)_{10.5\,K}$ from 2.0 to 10.4 K. The dotted lines are the best fits of data to the modified BTK model with two gaps plus intergrain Josephson effects (see the main text for details). **d,** Magnetic field evolution of $(dI/dV)/(dI/dV)_{8.5\,T}$ from 0.0 to 7.5 T at a fixed temperature of 5.0 K. All the curves are displayed with an offset for clarity. **e,** Contour plot of $(dI/dV)/(dI/dV)_{10.5\,K}$ of temperature ($T$) vs. bias voltage ($V$), where green (red) represents larger (smaller) differential conductance. **f,** Contour plot of $(dI/dV)/(dI/dV)_{8.5\,T}$ of field ($\mu_0 H$) vs. bias voltage ($V$) at 5.0 K.

**Fig. 4: Temperature and magnetic field dependence of the superconducting gap of Y.**

**a,** Temperature dependence of the two superconducting (SC) energy gaps obtained from different contacts P1 (open symbols) and P2 (solid symbols) at 48.6 GPa. The black and red lines are the best fits using the two-band $s$-wave BCS framework (see the main text for details). The reduced SC energy gaps for contacts P1 and P2 against reduced temperature are plotted in the inset. **b,** Magnetic field dependence of the two SC energy gaps from P1 at 48.6 GPa and a fixed temperature of 5.0 K. The short-dashed lines are the estimated power-law behavior, $\Delta(H, 5.0\,K) = \Delta(0, 5.0\,K)\cdot(1-H/H_{c2})^{1/2}$. The reduced energy gaps as a function of the reduced field are plotted in the inset.

Figures

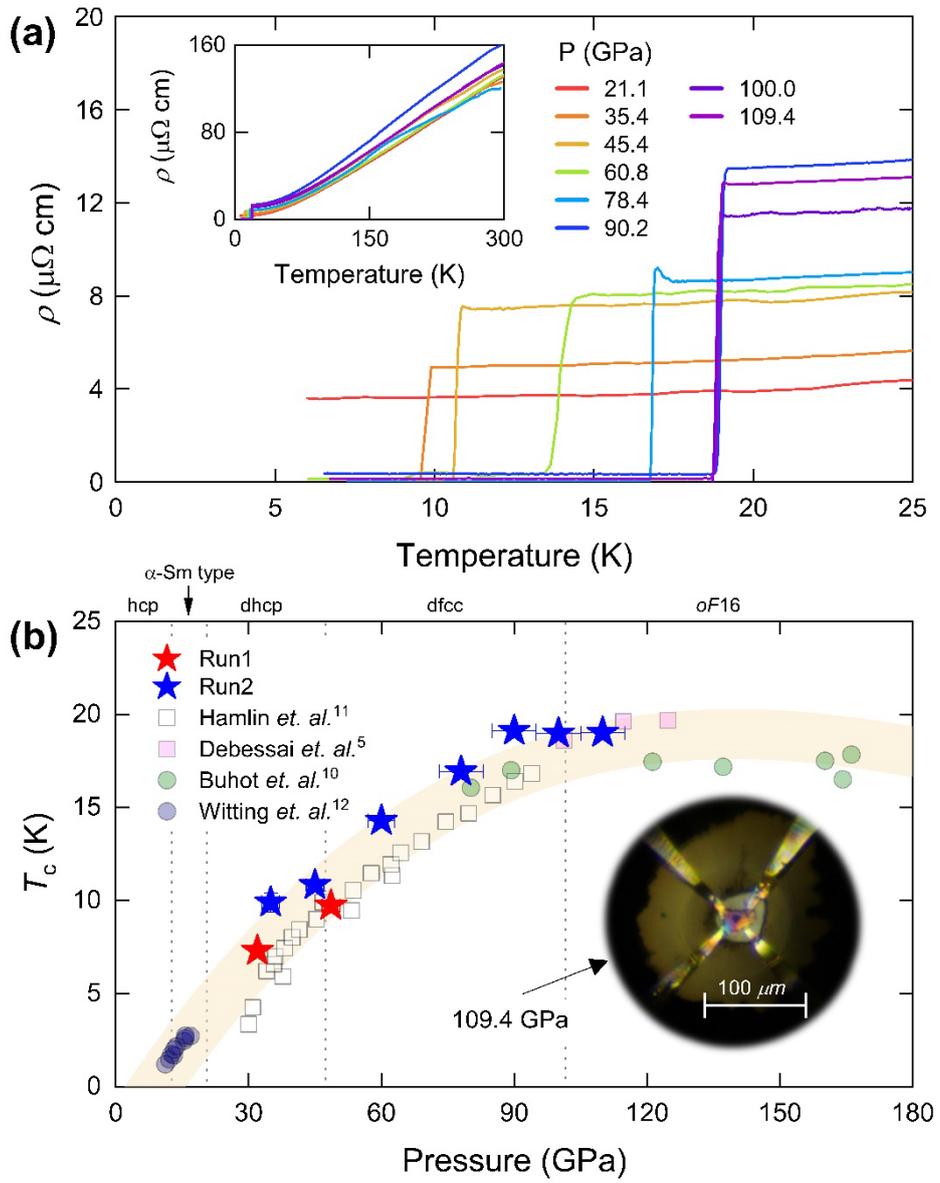

Figure 1

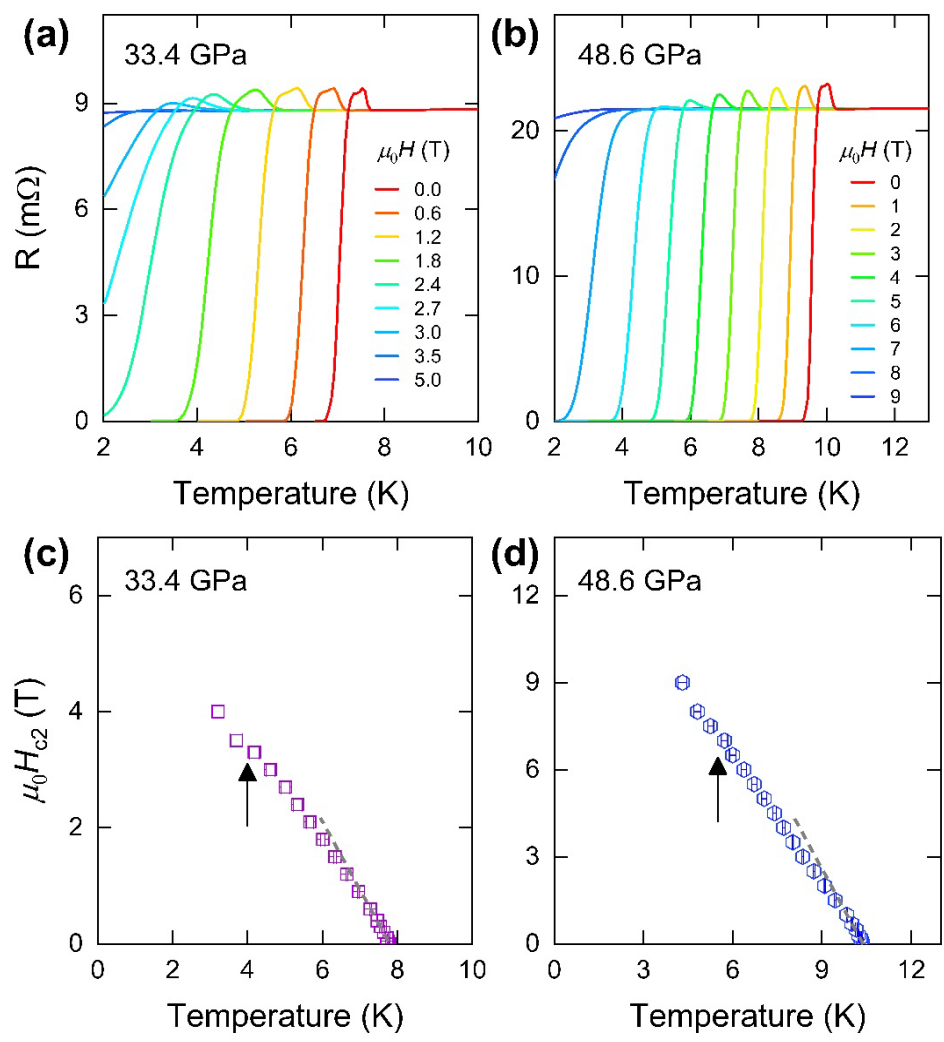

Figure2

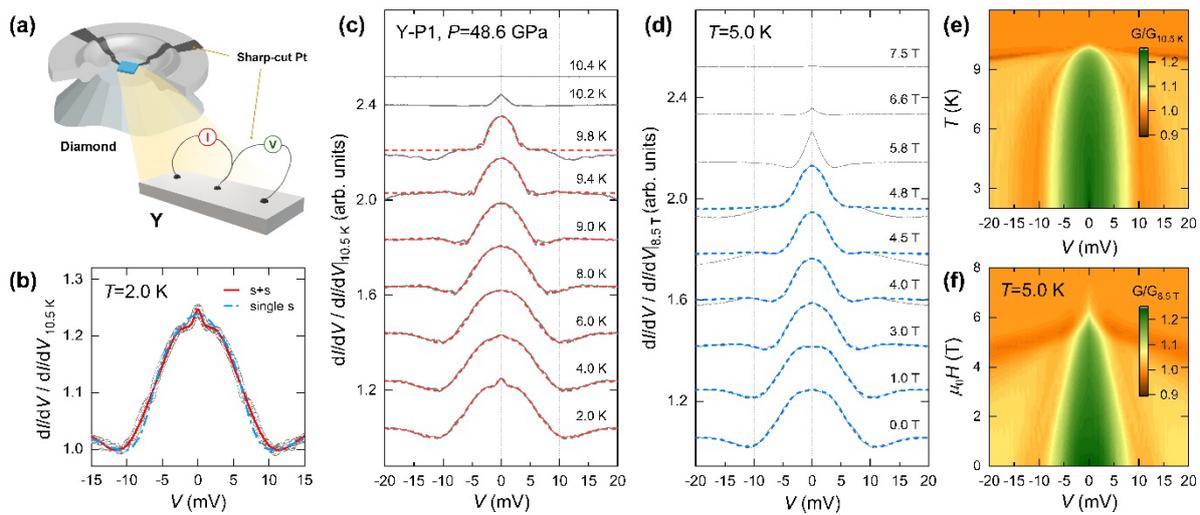

Figure 3

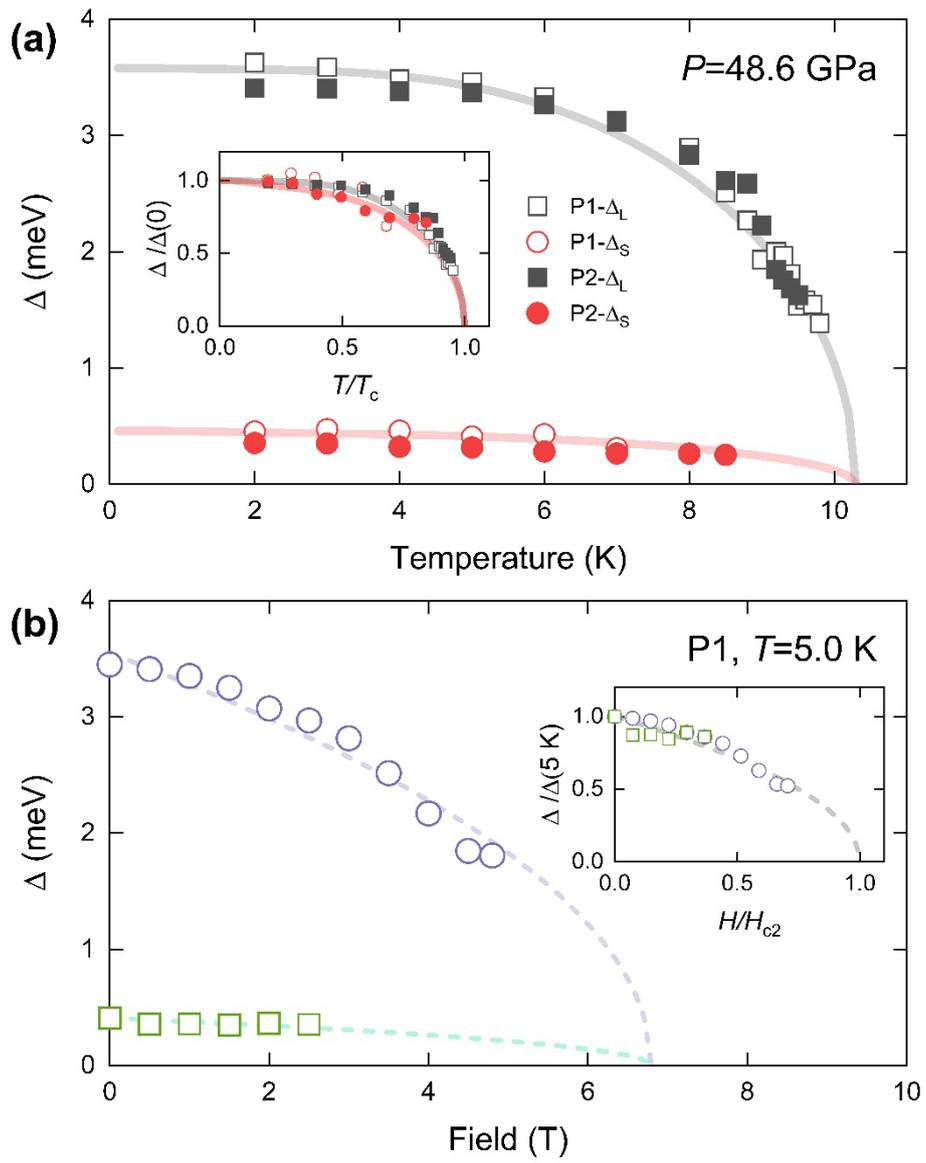

Figure 4